\documentclass[aps,twocolumn,showpacs,prl]{revtex4}
\usepackage{graphicx}
\usepackage{dcolumn}
\usepackage{amsmath}
\begin{document}
\title{Bichromatic Microwave Photoresistance of Two-Dimensional Electron System}
\author{M. A. Zudov$^{1,2}$}
\author{R. R. Du$^{1,3}$}
\affiliation{
$^1$Department of Physics, University of Utah, Salt Lake City, Utah 84112\\
$^2$School of Physics and Astronomy, University of Minnesota, Minneapolis, Minnesota 55455\\
$^3$Department of Physics and Astronomy, Rice University, Houston, Texas 77005
}
\author{L. N. Pfeiffer}
\author{K. W. West}
\affiliation{Bell Laboratories, Lucent Technologies, Murray Hill, New Jersey 07974}
\begin{abstract}
We explore experimentally bichromatic (frequencies $\omega _1$ and $\omega _2$) photoresistance of a two-di\-men\-si\-o\-nal electron system in the regimes of microwave-induced resistance oscillations and zero-resistance states.
We find bichromatic resistance to be well described by a superposition of $\omega _1$ and $\omega _2$ components, provided that both monochromatic resistances are positive.
This relation holds even when the oscillation amplitudes are small and one could expect additive contributions from monochromatic photoresistances.
In contrast, whenever a zero-resistance state is formed by one of the frequencies, such superposition relation breaks down and the bichromatic resistance is strongly suppressed.
\end{abstract}
\pacs{73.40.-c, 73.43.-f, 73.21.-b}
\maketitle
 
Microwave-induced resistance oscillations (MIRO) \cite{zudov:201311,ye:2193} and zero-resistance states (ZRS) \cite{mani:646, zudov:046807} emerging from the MIRO minima in ultraclean two-dimensional electron system (2DES), continue to attract attention of both experimentalists and theorists \cite{note1}. 
The resistance of the 2DES under microwave (MW) excitation of frequency $\omega$ can be expressed as $R^\omega = R^0 + \Delta R^\omega$, where $R^0$ is the dark value of resistance and $\Delta R^\omega$ is the radiation-induced change, or photoresistance (PR).
MIRO appear in experiment because $\Delta R^\omega$ oscillates wirh magnetic field, $B$, assuming both positive and negative values.
Giant $R^\omega$ maxima ($+$) and minima ($-$) occur at $B^{\pm}_j$, given by $\varepsilon^{\pm}_j= j\mp\phi^{\pm}_j$, where $\varepsilon=\omega/\omega_C$, $j=1,2,3,...$, $0<\phi \le 1/4$, and $\omega_C=eB/m^*$ is the cyclotron frequency \cite{zudov:041304}.
The majority of theoretical efforts were aimed at identifying microscopic mechanisms which could account for the generic properties of MIRO.
Early theories employed MW-induced short-range scattering \citep{durst:086803, lei:226805, vavilov:035303} (see also  \citep{ryzhii:2078, ryzhii:1299}), but it was subsequently argued that MW-induced oscillations in the non-equilibrium electron distribution function  \citep{dmitriev:165305, dmitriev:115316} (see also  \cite{dorozhkin:577}) are the leading cause of MIRO.
Other proposed mechanisms are based on photon-assisted effects  \citep{shi:086801}, gap formation  \citep{rivera:075314}, plasma oscillations  \citep{mikhailov:165311}, and non-parabolicity \citep{koulakov:115324}. 

In ultra-high quality samples $|\Delta R^\omega|$ easily exceeds $R^0$ and, while one could expect negative resistance at the MIRO minima, experimental $R^\omega$ usually saturates at zero over a finite range of $B$ giving rise to ZRS.
According to a general macroscopic argument  \citep{andreev:056803}, this happens because 2DES cannot sustain a negative resistance state.
On the other hand, zero-resistance state can be accommodated by forming current domains characterized by a unique magnitude of the local current density at which the voltage drop vanishes.
Although there exist other interesting proposals that do not invoke negative resistance, such as those based on sliding charge-density waves \cite{phillips:233}, or quantum coherence effects \cite{lee:115336}, the domain model remains the most popular. 
Experimentally, it is established that current patterns in MW-irradiated 2DES hint at anomalous flow  \citep{willett:026804} and could be otherwise very complicated  \citep{zudov:aps}, but the direct spatial imaging of current domains remains a subject of future experiments. 
Among other outstanding issues are activated temperature dependence with large energy gaps  \citep{mani:646, zudov:046807, willett:026804}, hysteresis loops in magnetic field  \citep{willett:026804} and power \citep{zudov:aps} sweeps, suppression of MIRO/ZRS by modest in-plane magnetic fields  \citep{yang:up} (see, however, \cite{mani:075327}), immunity of MIRO/ZRS to the sense of circular polarization of MWs  \citep{smet:116804}, and relevance of multi-photon processes in the ZRS formation  \cite{zudov:041303}. 
As none of these findings can be readily accomodated by mainstream theories, further experimental developments are needed to pinpoint the origin of these fascinating phenomena.

In this Letter we report on magnetotransport measurements in an ultraclean 2DES exposed to {\em bichromatic} MW radiation from two monochromatic sources of frequencies $\omega_1$ and $\omega_2$.
We find that 
1) whenever both monochromatic resistances are positive, bichromatic resistance is well described by a superposition of $\omega_1$ and $\omega_2$ resistances;
and 2) this relation fails whenever a ZRS is formed in one of the monochromatic resistances.
Here, bichromatic resistance is dramatically reduced with respect to what is anticipated from a superposition.
While the specific nature of both observations remains a subject of future experimental and theoretical work, the latter appears to support the concept of absolute negative resistance theoretically linked to monochromatic ZRS.
As will be shown, there also exists a more subtle relation between bichromatic and monochromatic resistances than 1) and 2).

Our sample was cleaved from a symmetrically doped Al$_{0.24}$Ga$_{0.76}$As/ GaAs/Al$_{0.24}$Ga$_{0.76}$As 300\,\AA-wide quantum well grown by molecular beam epitaxy.
After illumination with visible light, low-temperature electron mobility, $\mu$ and density, $n_e$ were $\sim$\,$2 \times 10^7$ cm$^2$/Vs and $3.6 \times 10^{11}$ cm$^{-2}$, respectively.
Basic experimental details are described in Refs. \cite{zudov:201311,zudov:046807,zudov:041304}.
Bichromatic radiation is obtained by combining MWs from two Gunn diodes tuned to distinct frequencies, using a hybrid ``T'' mounted on the top flange of the WR-28 waveguide used to carry radiation down to the sample.
While similar results were obtained for several sets of MW frequencies, all the data reported here were recorded with $f_1=\omega_1/2\pi=31$ GHz and/or $f_2=\omega_2/2\pi=47$ GHz.
Since in the ZRS regime resistance peaks occur close to cyclotron resonance harmonics  \cite{zudov:046807,zudov:041304}, such choice of frequencies ($f_2/f_1\approx$\,1.5) allows overlaps of a variety of monochromatic features.
Indeed, all possible situations, i.e., overlaps of maxima ($\uparrow$) at $B^{\omega_1+}_{2(4)}\approx B^{\omega_2+}_{3(6)}$, of minima ($\downarrow$) at $B^{\omega_1-}_{2(3)}\approx B^{\omega_2-}_{3(5)}$, and of a maximum with a minimum ($\updownarrow$) at $B^{\omega_1-}_{1}\approx B^{\omega_2+}_{2}$ and at $B^{\omega_1+}_{3}\approx B^{\omega_2-}_{4}$, are experimentally accessible with this choice of frequencies.

\begin{figure}[t]
\includegraphics{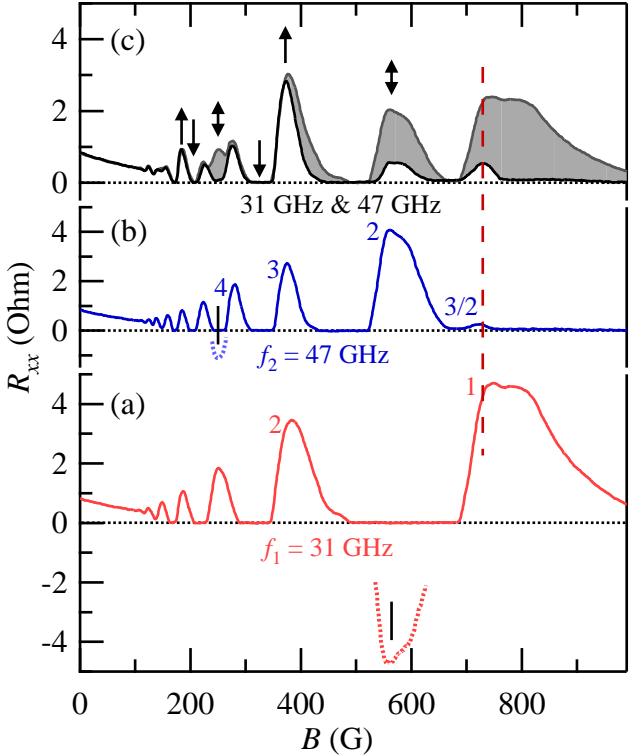}
\caption{[color online]
Panels (a) and (b) show magnetoresistance data $R^{\omega_1}$ and $R^{\omega_2}$ under monochromatic radiation of frequencies $f_1=31$ GHz and $f_2=47$ GHz, respectively.
Numbers label harmonics of cyclotron resonance.
In panel (c) solid line represents experimental bichromatic magnetoresistance ($f_1$ and $f_2$).
Maximum-maximum, minimum-minimum and maximum-minimum overlaps are marked by $\uparrow$, $\downarrow$, and $\updownarrow$, respectively.
The upper boundary of the shaded area represents the average of monochromatic resistances $R^{\omega_1}$ and $R^{\omega_2}$ presented in (a) and (b) (see text).
Dotted line in (a) represents reconstructed negative resistance as described in the text.
}
\label{tc1}
\end{figure}
In Fig.\,\ref{tc1} we present an overview of our experimental data showing magnetoresistance acquired under: (a) $\omega_1$-radiation, (b) $\omega_2$-radiation, and (c) bichromatic $\omega_1,\omega_2$-radiation, all plotted using the same scaling for easy comparison.
Monochromatic traces attest to exceptional quality of our specimen; at both frequencies ZRS persist down to $B\approx$\,175 G, yielding four(six) ZRS for $\omega_1$($\omega_2$).
Of primary interest is the resistance under bichromatic excitation, shown by a black solid line in (c), and its relation to monochromatic traces.
First, we notice that the amplitude of the bichromatic PR is rather irregular as a function of $B$.
Second, the bichromatic signal is generally weaker unless both frequencies induce positive PR of similar magnitude, e.g., when both monochromatic resonant conditions for the maxima are satisfied.
Indeed, when monochromatic peaks overlap, e.g. at $B^{\omega_1+}_{2(4)} \approx B^{\omega_2+}_{3(6)}$, the bichromatic response mimics the monochromatic responses;
the peaks at 380 and 185 G, marked by $\uparrow$ in (c), are of about the same height as corresponding monochromatic peaks in (a) and (b).
Similarly, overlapped monochromatic ZRS survive in bichromatic resistance, e.g. at $B^{\omega_1-}_{2(3)}\approx B^{\omega_2-}_{3(5)}$, as marked by $\downarrow$.
On the other hand, when a peak overlaps with ZRS, the contributions from different frequencies tend to cancel each other and the fate of bichromatic resistance cannot be readily predicted.
For example, bichromatic trace shows a deep minimum at $B^{\omega_1+}_{3}\approx B^{\omega_2-}_{4}\approx 250$ G (left $\updownarrow$), but an overlap at $B^{\omega_1-}_{1}\approx B^{\omega_2+}_{2} \approx 570$ G results in a peak (right $\updownarrow$).

In order to get a deeper understanding of the bichromatic PR, we turn our attention to the lower magnetic field regime, where monochromatic PR exhibits MIRO. 
In Fig.\,2 we replot monochromaric $R^{\omega_1}$(dashed line), $R^{\omega_2}$(dotted line), and bichromatic $R^{\omega_1\omega_2}$(solid line) as a function of $1/B$ and observe that the bichromatic resistance nowhere exceeds monochromatic resistances.
At $B^{-1}\approx$\,5.4 kG$^{-1}$ we confirm that the fourth $\omega_1$ and the sixth $\omega_2$ maxima produce comparable bichromatic peak.
More surprisingly, this behavior persists down to much lower magnetic fields, until the MIRO disappear. 
While one can anticipate additive contributions from monochromatic PRs (more so when oscillations are weak), bichromatic resistance simply repeats nearly coinciding monochromatic peaks ($B^{\omega_1+}_{6} \approx B^{\omega_2+}_{9}$) marked by $\uparrow$ at $B^{-1} \approx 8$ kG$^{-1}$.
Moreover, below the onset of ZRS ($B^{-1}\gtrsim 6$ kG$^{-1}$), bichromatic resistance closely follows the average of monochromatic resistances, as further confirmed by readily identified common crossing points (open circles).
All these observations suggest that averaging seems to hold well in the MIRO regime, regardless of the PR sign.

We now return to the examination of the data in the ZRS regime presented in Fig.\,1, where we have already observed the averaging for the peaks at $B^{\omega_1+}_{2(4)} \approx B^{\omega_2+}_{3(6)}$ ($\uparrow$).
Averaging also obviously holds when monochromatic ZRS are overlapped, e.g., at $B^{\omega_1-}_{2(3)}\approx B^{\omega_2-}_{3(5)}$ ($\downarrow$).
To check how averaging holds over the entire field range, we compare the average of experimental traces (a) and (b) with the bichromatic trace in (c).
The difference between the two is shown by the shaded area in Fig.\,1(c), where the upper boundary is the average of the experimental traces (a) and (b).
We immediately observe that whenever a ZRS is overlapped with a peak, e.g. at $B^{\omega_1+}_{3}\approx B^{\omega_2-}_{4}$ or $B^{\omega_1-}_{1}\approx B^{\omega_2+}_{2}$ (marked by $\updownarrow$), the bichromatic response is suppressed below the average of monochromatic responses. 
While this suppression is always present in the ZRS regime, it becomes progressively smaller at lower magnetic fields and disappears completely in the MIRO regime.
As evidenced by a partial overlap of $B^{\omega_1-}_{3} \approx B^{\omega_2+}_{5}$ marked by $\updownarrow$ at $B^{-1}\approx 0.45$ kG$^{-1}$ in Fig.\,2, the edge of a rather weak ZRS gives rise to only a slight suppression of the bichromatic PR below the average.

\begin{figure}[t]
\includegraphics{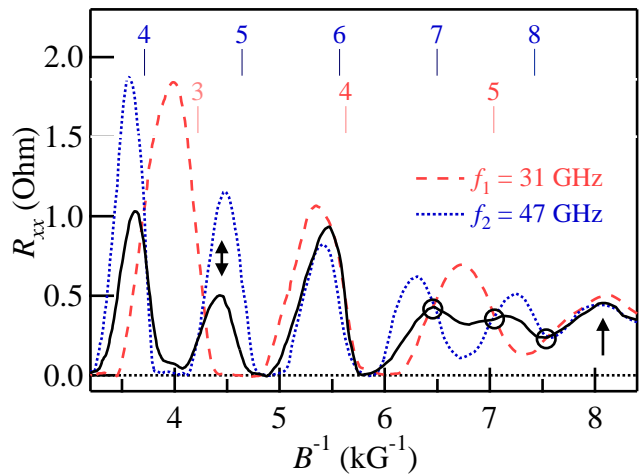}
\caption
{[color online]
Monochromaric $R^{\omega_1}$(dashed line), $R^{\omega_2}$(dotted line), and bichromatic $R^{\omega_1\omega_2}$(solid line) resistance versus inverse magnetic field.
Upper (lower) set of numbers labels $\omega_2$ ($\omega_1$) CR harmonics.
}
\label{tc2}
\end{figure}
\begin{figure}[t]
\includegraphics{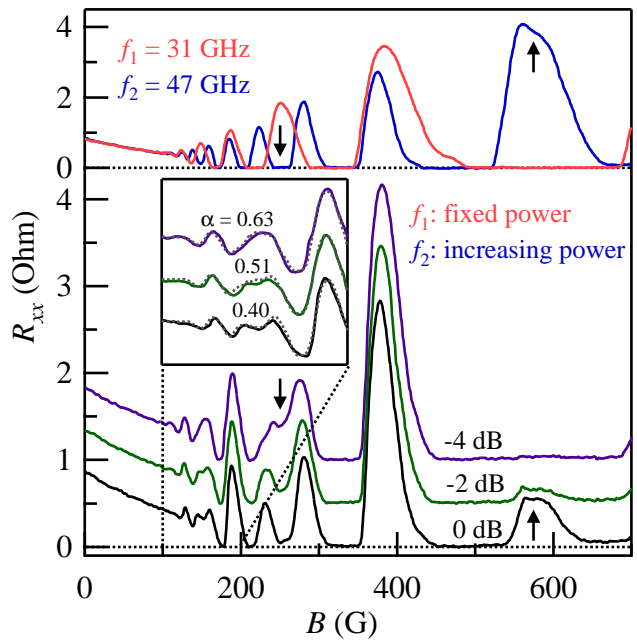}
\caption
{[color online]
Top two traces are monochromatic magnetoresistances $R^{\omega_1}$ and $R^{\omega_2}$.
Vertical arrows mark the positions of $B^{\omega_2+}_{2}\approx B^{\omega_1-}_{1}$ ($\uparrow$) and
$B^{\omega_2-}_{4}\approx B^{\omega_1+}_{3}$ ($\downarrow$).
Bottom three traces show bichromatic magnetoresistance at various intensities of the $\omega_2$-source as marked by attenuation.
In the inset solid lines duplicate the magnetoresistance in the MIRO regime ($100$ G $<B<200$ G) and dotted lines are calculated as described in the text.
}
\label{tc3}
\end{figure}

While one might be tempted to conclude that bichromatic resistance is well described by the average, whenever monochromatic resistances do not exhibit ZRS, more thorough examination of the data in Fig.\,1 reveals a special situation which calls for a different interpretation. 
More specifically, at $B\approx 730$\,G, (see dashed vertical line in Fig.\,1) $R^{\omega_2}$ exhibits a small peak of multiphoton origin  \cite{zudov:041304, zudov:041303} which overlaps with the fundamental $\omega_1$ peak.
ZRS is observed in neither (a) nor (b), but bichromatic resistance is suppressed considerably below the average.
While this observation deserves further studies, it can be related to the contribution of frequency mixing processes which might alter the resonance condition and hence the sign of the PR \cite{zudov:041303}.
In particular, a two-photon $\omega_1+\omega_2$ process is expected to give rise to a negative PR.
If it prevails over a two-photon $\omega_2+\omega_2$ process, responsible for a small peak in monochromatic PR in Fig.\,1(b), bichromatic PR will be reduced below the average value. 

It is interesting to examine bichromatic resistance for various mixtures of $\omega_1$ and $\omega_2$ MW fields.
In this experiment we keep the intensity of the $\omega_1$ source constant and vary the intensity of the $\omega_2$ source.
In Fig.\,3 (bottom) we plot bichromatic resistance for selected intensities of the $\omega_2$ source, as marked by attenuation (-4, -2, and 0 dB).
The traces are vertically offset for clarity and we also include monochromatic traces at the top of the figure for reference.
We observe that whenever two ZRS or two peaks overlap, bichromatic resistance has little dependence on the intensity of the $\omega_2$ source.
On the contrary, when a peak is overlapped with ZRS, bichromatic PR exhibits substantial changes.
For example, initially well developed ZRS associated with $B^{\omega_1-}_{1}$ (marked by $\uparrow$), is partially destroyed with increasing $\omega_2$ power; here, $B^{\omega_2+}_{2}$ peak induces a reentrant ZRS structure about $B^{\omega_1-}_{1}$.
Similarly, $\downarrow$ marks the development of the ZRS about $B^{\omega_2-}_{4} $at the position of the $B^{\omega_1+}_{2}$ peak with increasing $\omega_2$ power.

Now we try to offer some interpretation of our data within a framework of the negative resistance and its instability.
In the regime of linear response conductivity, one can expect additivity $\Delta R^{\omega_1\omega_2}=\Delta R^{\omega_1} + \Delta R^{\omega_2}$, where monochromatic PRs commensurate with the intensity of the corresponding frequency component.
Observation of the ZRS typically requires intense radiation and the PR tends to saturate \cite{zudov:041303}.
As far as we know, this regime is yet to receive due theoretical treatment which would allow for comparison with experiment, even for the monochromatic case.
Experimentally, we have observed that under these conditions the bichromatic PR does not exceed monochromatic PRs, i.e., $|\Delta R^{\omega_1\omega_2}|<|\Delta R^{\omega_1}|,|\Delta R^{\omega_2}|$.
Moreover, we have observed that monochromatic frequencies produce traces of similar oscillation amplitude and bichromatic PR closely follows the average of monochromatic resistances.
The simplest relation between $R^{\omega_1}$, $R^{\omega_2}$ and $R^{\omega_1\omega_2}$ which properly accounts for saturation is obtained from a linear partition between two monochromatic contributions, $\Delta R^{\omega_1\omega_2}=\alpha\cdot\Delta R^{\omega_1} + (1-\alpha)\cdot\Delta R^{\omega_2}$, which clearly holds when $\omega_1=\omega_2$. 
Such empirical relation would mean that second MW field causes the redistribution of scattering events and each monochromatic contribution depends on the relative intensity of the corresponding frequency component.
Obviously, this relation directly translates to the superposition for the total resistance and experimental bichromatic PR becomes:
\begin{equation}
R^{\omega_1\omega_2}_{\exp}=\max \left \{\alpha \cdot R^{\omega_1} + (1-\alpha) \cdot R^{\omega_2}, 0 \right\}.
\label{dc}
\end{equation}
In our experiments monochromatic PRs are saturated at about the same strength, which implies $\alpha \sim 1/2$ and explains observed average response.
If we assume $\alpha = (1+p)^{-1}$, where $p=P_2/P_1 \propto P_2$ and $P_1$($P_2$) is the intensity of the $\omega_1$($\omega_2$) source, we can check the superposition relation against our data in Fig.\,3, as we vary relative intensity via $P_2$.
A good fit at 0 dB is achieved with $\alpha=0.4$ which translates to $p_0=1.5$. 
Using known attenuations of the $\omega_2$ source we find $\alpha(-2)=(1+p_0\cdot 10^{-0.2})^{-1} \approx 0.51$ and $\alpha(-4)= (1+p_0\cdot 10^{-0.4})^{-1}\approx 0.63$ for the other two traces and then compute bichromatic resistance using Eq.\,(\ref{dc}). 
The results of these calculations, presented in the inset of Fig.\,3 by dotted lines and marked with $\alpha$, demonstrate excellent agreement with experiment.

Eq.\,(\ref{dc}) not only provides good description of our experimental data (apart from ZRS and multi-photon features), but it also allows us to relate the observed suppression below the average to the negative resistance associated with the monochromatic ZRS.
This negative resistance enters the right-hand side of Eq.\,(\ref{dc}) and if it does not exceed the positive term by absolute value, measured bichromatic resistance will still be positive.
For example, the peak at $B^{\omega_2+}_{2}$ ($\updownarrow$) survives in bichromatic resistance, albeit with a dramatically reduced amplitude.
This reduction can be qualitatively attributed to the negative resistance associated with the ZRS at $B^{\omega_1-}_{1}$.
Bearing the inherent oversimplification of Eq.\,(\ref{dc}), one can also obtain a quantitative estimate of the negative resistance over a finite range of magnetic field, limited by positive bichromatic resistance, as, e.g., $R^{\omega_1(\omega_2)} = 2R^{\omega_1\omega_2}_{\exp} - R^{\omega_2(\omega_1)}$.
Such reconstruction was performed for both frequencies at about $B^{\omega_1-}_{1}$ and $B^{\omega_2-}_{4}$ using $\alpha=0.4$
The results, shown by dotted lines and marked by $|$ in Fig.\,\ref{tc1}(a) and (b), appear in reasonable agreement with monochromatic MIRO envelopes.
Bichromatic measurements, therefore, allow, in principle, to access negative resistance which is masked by instability in monochromatic experiments.

In summary, we have studied bichromatic PR of a high-quality 2DES in the regimes of MIRO and ZRS.
In the MIRO regime, or whenever both monochromatic resistances are positive, we have found that bichromatic resistance is close to a simple superposition of the monochromatic resistances.
This behavior persists down to very low magnetic fields until the PR disappears.
In the ZRS regime, bichromatic resistance was found to be substantially lower than the superposition of the two components.
This observation can be viewed as a qualitative evidence of absolute negative resistance associated with monochromatic ZRS, which then enters empirical superposition relation.
It will be interesting to see if theoretical support for such a relation under strong bichromatic radiation becomes available.
The bichromatic probe can also be potentially useful to address other interesting issues, such as coherence effects in electronic transitions with multiple channels \cite{lee:115336} and parametric resonances \cite{joas:235302}.

We would like to thank M.E. Raikh, F. von Oppen, E.E. Kolomeitsev, and L.I. Glazman for helpful discussions.
This work was supported by DARPA QuIST and NSF DMR-0408671.



\end{document}